\documentstyle[aps,psfig]{revtex}
\begin{document}
\title{ Distributions of Singular Values for Some Random Matrices} 
\author{A.M.Sengupta}
\address{Serin Physics Laboratory, Rutgers University,Piscataway, NJ
08854} 
\address{and}
\author{P.P. Mitra}
\address{Bell Laboratories, Lucent Technologies, 600 Mountain Avenue,
Murray Hill, NJ07974}
\maketitle
\begin{abstract}
The Singular Value Decomposition is a matrix decomposition technique
widely used in the analysis of multivariate data,  such as complex
space-time images obtained in both physical and biological systems. In 
this paper, we examine the distribution of Singular Values of low
rank matrices corrupted by additive noise. Past studies have been
limited to uniform uncorrelated noise. Using diagrammatic and saddle
point integration
techniques, we extend these results to heterogeneous and
correlated 
noise sources. We also provide perturbative estimates of error bars on
the reconstructed low rank matrix obtained by truncating a Singular
Value 
Decomposition.  
\end{abstract}

In analysing large, multivariate data, certain quantities naturally
arise 
that are in some sense `self averaging', namely in the large size limit,
a 
single data set can comprise a statistical ensemble for the quantity in 
question. One such quantity, the singular value distribution of a data
matrix, 
is the subject of this paper. The Singular Value Decomposition (SVD) is
a 
representation of a general matrix of fundamental importance in linear
algebra 
that is widely used to generate canonical representations of
multivariate data. It is equivalent to Principal Component Analysis in
multivariate statistics, 
but in addition is used to generate low dimensional representations for
complex
multi-dimensional time series.  One example is to generate effective low 
dimensional representations of high dimensional dynamical systems. 
Another 
example of curent interest is to de-noise and compress dynamic imaging
data, 
in particular in the case of direct or indirect images of neuronal
activity.
        
For most of the above applications  it is important to understand 
the properties of an SVD of a matrix whose entries show some degree of
random 
fluctuations.  This has  been achieved to an extent in multivariate
statistics,
where the sampling distributions of quantities associated with the PCA
are 
computed; however, the computations involved are difficult and exact 
distributional results are limited. 

In this paper, we use diagrammatic and saddle point integration
techniques
to 
obtain the densities of singular values of matrices whose entries have
varying 
degrees of randomness.  In particular, we study the problem in the
asymptotic 
limit of large matrix size; this limit is well justified in realistic
cases as will be described below.  The density of SV's has been obtained
before,
with other techniques, for  matrices with each entry independently
distributed 
normally with indentical variances \cite{Hua,Denby}.  We are able 
to obtain distributions for 
some more general cases where the variances are not equal and/or
correlations 
are present between matrix entries.
Our results have 
implications towards isolating random components from image time series.
Also, 
these results help in understanding the effects of 
truncating the SV spectrum at a given point, a technique that is widely
applied
to remove noise from data.

The SVD of an arbitrary (in general complex) 
$p\times q$ matrix ($p \geq q$) ${\cal M}$ is given by  
${\cal M} = U \Lambda V^{\dagger}$, where the $p\times q$ matrix $U$ has 
orthonormal rows, the $q\times q$ matrix $\Lambda$   is diagonal with
real, 
non-negative entries and the $q \times q$ matrix $V$ is unitary. 
  Note that the matrices ${\cal M} {\cal M}^{\dagger} = U \Lambda^2
U^{\dagger}$
and ${\cal M}^{\dagger} {\cal M} = V^{\dagger} \Lambda^2  V$
 are hermitian, with eigenvalues corresponding to 
 the diagonal entries of  $\Lambda^2$  and U and V 
the corresponding matrices of eigenvectors.  Consider the special case
of 
space-time data
$I({\bf x}, t)$. 
The SVD of such data is given by 
\begin{equation}
\label{svd}
I({\bf x} , t) = \sum_n \lambda_n I_n({\bf x}) a_n(t)
\end{equation} 
where $I_n({\bf x})$  are the eigenmodes of the spatial ``correlation"
matrix
$C({\bf x}, {\bf x}^{\prime}) = \sum_t I({\bf x},t) I({\bf
x}^{\prime},t) $,
and similarly $a_n(t)$ are the eigenmodes of the ``temporal correlation
function"
$C(t, t^{\prime}) = \sum_t I({\bf x},t) I({\bf x},t^{\prime}) $.
If one considered the sequence of images as randomly chosen from an
ensemble 
of spatial images, then $C({\bf x}, {\bf x}^{\prime})$  would converge
to the 
ensemble spatial correlation function in the limit of long times.  If in 
addition the ensemble had space translational invariance then the
 eigenmodes   $I_n({\bf x})$ would be plane waves $e^{i{\bf k}\cdot {\bf
x}}$,
 the mode number ``n" would correspond to wavevectors and the singular
values 
would correspond to the spatial structure factor $S({\bf k})$.
In general, the image ensemble in question will not have translational 
invariance; however the SVD will then provide a basis set analogous to
wave 
vectors.  In physics one normally encounters the structure factors
$S({\bf k})$ that decay with wave vector, and in the more general case,
the singular value 
spectrum, organized in descending order, will show a decay indicating
the 
structure in the data.

We consider the case of a $p\times q$ matrix ${\cal M} = {\cal M}_0 + N$
   where  ${\cal M}_0$ is fixed and the entries of $N$ are normally
distributed
 with zero mean. We consider below several cases of normal distributions
for 
entries of $N$, including cases where there are correlations between
entries 
of $N$.   ${\cal M}_0$ may be thought of as the desired or underlying
signal;  
for an SVD to be useful, ${\cal M}_0$  should effectively have a low
rank 
structure.

It is convenient to work in terms of the resolvent
\begin{equation}   
{\cal G}(z) = Tr[(z -  {\cal M}^{\dagger} {\cal M})^{-1}]  
= \sum_n {1\over z-\lambda_n^2}
\end{equation}
where  ${\cal G}(z)$  is a complex function. The density of SV's
 is given by
\begin{equation}
\rho(\lambda) = \sum_n \delta(\lambda-\lambda_n) = {2 \lambda \over \pi}
\lim _{\epsilon \rightarrow 0} Im \Bigl[ {\cal G}(\lambda^2-i\epsilon) - 
{\cal G}(\lambda^2-i\epsilon) \Bigr]
\end{equation}
We proceed by finding a closed equation for the average of ${\cal G}(z)$
in the
limit where $p,q \rightarrow \infty$ and $\sigma_i \rightarrow 0$ such
that the
variance of a matrix entry times q, $\sigma^2 q $ converges to a finite
limit 
$\tilde{\sigma}^2$ and $p/q$ is held fixed.  
The density of states is expected to be a self averaging quantity. Thus
although we compute results by averaging over the ensemble, we expect to
be 
able to apply our results to the SVD of individual data matrices.

To illustrate the method, consider the simplest case, where each element
of the
matrix is {\it i.i.d.} normally as ${\cal N}(0,\sigma)$. Since 
\begin{equation}
{\cal G}(z) = \partial_z \ln \det (z - {\cal M}^{\dagger} {\cal M})
\label{gz}
\end{equation}
 the average of the resolvent over the 
probability distribution of the matrix ${\cal M}$ can be obtained from 
$\langle \ln \det (z - {\cal M}^{\dagger}{\cal M}) \rangle$, which in
turn
 may be computed using 
replicas. We introduce $n$ replicas of $q$ dimensional real vectors 
$X_{\alpha} = (x_{\alpha 1}, .., x_{\alpha q}) (\alpha = 1,..,n)$.
Consider the following identity 
\begin{equation}
Z_n = \int \bigl[ \Pi_{\alpha =1}^n \Pi_{a =1}^q \bigr] 
\Bigl\langle exp\bigl( -{q\over 2} \sum_{\alpha = 1}^n X_{\alpha}^T
 (z - {\cal M}^{\dagger}{\cal M})
X_{\alpha} \bigr) \Bigr\rangle = 
({2\pi \over q})^{nq\over 2} 
\langle \bigl[\det (z - {\cal M}^{\dagger}{\cal M})\bigr]^{-{n\over 2}}
\rangle
\end{equation} 
One obtains the desired quantity from the above by taking $n\rightarrow
0$. 
Before 
computing the expectation over ${\cal M} $ in $Z_n$, we decouple the
term 
$X_{\alpha}^T {\cal M}^{\dagger}{\cal M} X_{\alpha}$ using the 
Hubbard-Stratanovich 
transformation by introducing another $n$ vectors 
$Y_{\alpha} = (y_{\alpha 1}, .., y_{\alpha p}) (\alpha = 1,..,n)$ which
are $p$ 
dimensional. 
Performing the average over ${\cal M}$, we obtain 
\begin{equation}
Z_n = ({q\over 2\pi}) ^{np\over 2} \int\bigl[ \prod dX dY \bigr]
\exp\Bigl(- {q\over 2} \bigl[ \sum_{\alpha} (z X_{\alpha}^T X_{\alpha} 
+ Y_{\alpha}^T Y_{\alpha})
-   \tilde{\sigma}^2 \sum_{\alpha \beta} X_{\beta}^T X_{\alpha}
Y_{\alpha}^T
 Y_{\beta}\bigr] \Bigr)
\end{equation}

Now, we decouple $\sum_{\alpha \beta} X_{\beta}^T X_{\alpha}
Y_{\alpha}^T
 Y_{\beta}$ by using two $n\times n$ matrices $[Q_{\alpha \beta}]$ and
$[R_{\alpha \beta}]$ as follows:
\begin{eqnarray*}
 Z_n = 2^{-n^2}({q\over 2\pi}) ^{{np\over 2}+ n^2} 
\int&&\bigl[ \prod dX dY dR dQ \bigr] \\
&&\exp\Bigl(- {q\over 2} \bigl[ \sum_{\alpha \beta} ( X_{\alpha}^T 
(z \delta_{\alpha \beta}
 -\tilde{\sigma}^2 Q_{\alpha \beta})X_{\beta}
 + Y_{\alpha}^T(\delta_{\alpha \beta}
 - i R_{\alpha \beta}) Y_{\beta}
+ i  Q_{\beta \alpha} 
R_{\alpha \beta})\bigr] \Bigr)
\end{eqnarray*}
The integral over $R$ enforces the condition $Q_{\alpha \beta}= 
Y_{\alpha} Y_{\beta}^T $, giving us back the previous expression.

At this stage, we are in a position to write $Z_n$ as an integral over
$Q$ and $R$ only, since we can do the integrals over $X$ and $Y$.
\begin{equation}
Z_n = 2^{-n^2}({q\over 2\pi}) ^{-{nq\over 2}+n^2} \int\bigl[ \prod dR dQ
\bigr]
\exp\Bigl(- {q\over 2} \bigl[\ln \det (z -\tilde{\sigma}^2 Q)
 + {p \over q} \ln \det (1 - i R)
+ i  tr (Q R)\bigr] \Bigr)
\end{equation}

Ideally one should take the $n \rightarrow 0$ limit first and  then let
 $q \rightarrow \infty$. In order to be able to perform analytical 
computations, we have to take the limit in the reverse order .
That this gives the correct answer is verified later by a direct
diagrammatic method.

When $p,q \rightarrow \infty $, with $p/q$ fixed, the integral is
dominated
 by some saddle point. We take the replica diagonal ansatz, consistent
with
all the symmetries.

\begin{equation}
Q_{\alpha \beta}=Q(z)\delta_{\alpha \beta}
\end{equation}
\begin{equation}
R_{\alpha \beta}=iR(z)\delta_{\alpha \beta}
\end{equation}

Then we have to minimize
\begin{equation}
S\bigl(Q(z),R(z)\bigr) =\ln (z -\tilde{\sigma}^2 Q(z))
 + {p \over q} \ln (1 - R(z))
+ Q(z) R(z)
\end{equation}
with respect to $Q(z)$ and $R(z)$. Hence the equations,

\begin{equation}
{1 \over z-\tilde{\sigma}^2 Q(z)}={1 \over { \tilde{\sigma}^2}}R(z)
\end{equation}
\begin{equation}
{p/q \over 1-R(z)}=Q(z)
\end{equation}

Using the fact that $Z_n \sim \exp \bigl[-nq S\bigl(Q(z),R(z)\bigr)
\bigr]$
and \ref{gz}, we get
\begin{equation}
G(z)=\langle {\cal G}(z) \rangle={q \over z-\tilde{\sigma}^2Q(z)}
\end{equation}
so that $ G(z)$ satisfies
\begin{equation}
 G(z)={q \over z-{p\tilde {\sigma}^2 \over q - \tilde {\sigma}^2 G(z) }}
\end{equation}

This equation can also be obtained from a direct diagrammatic
resummation
of $ \Bigl \langle Tr{1 \over z -  {\cal M}^{\dagger} {\cal M}} 
\Bigr\rangle$ expanded in powers of ${1 \over z}$. 
The diagrammatic representation of these terms are shown in Fig.1.
\begin{figure}[h!]
\centerline{\hbox{\psfig{figure=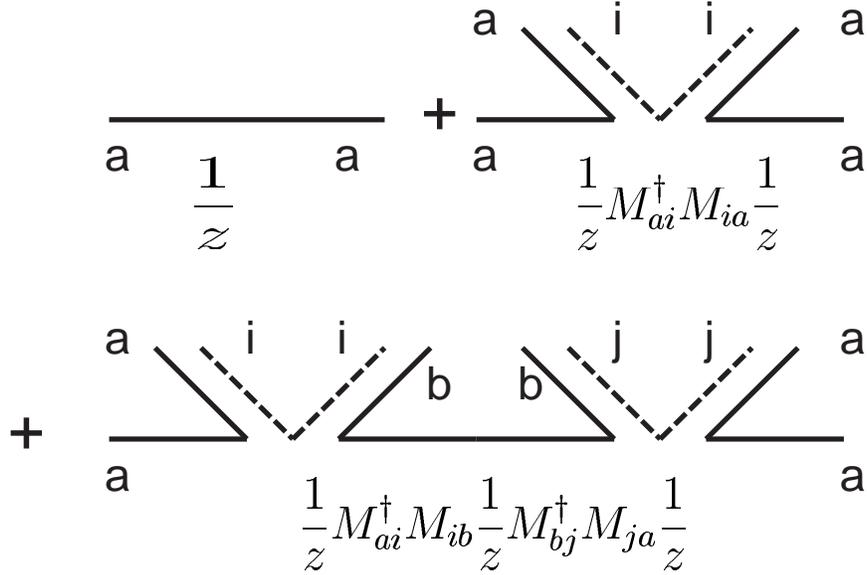,height=3in,width=4.5in}}}
\caption{Diagrammatic representation of successive terms
in the resolvent. }
\end{figure}
We use Wick's theorem to take the average over $M$ with
\begin{equation}
\langle M_{ia} M_{jb} \rangle = \sigma^2 \delta_{ij} \delta_{ab}
\end{equation}
We have to concentrate only on the one-particle irreducible graphs,
which give rise to the self-energy. The advantage of taking the large
$p$,$q$ limit is that we have  to consider only planar diagrams
\cite{'tHooft}.
The diagrams contributing to self-energy, in the large $p$,$q$ limit  
are shown schematically in the Fig.2.

\begin{figure}[h!]
\centerline{\hbox{\psfig{figure=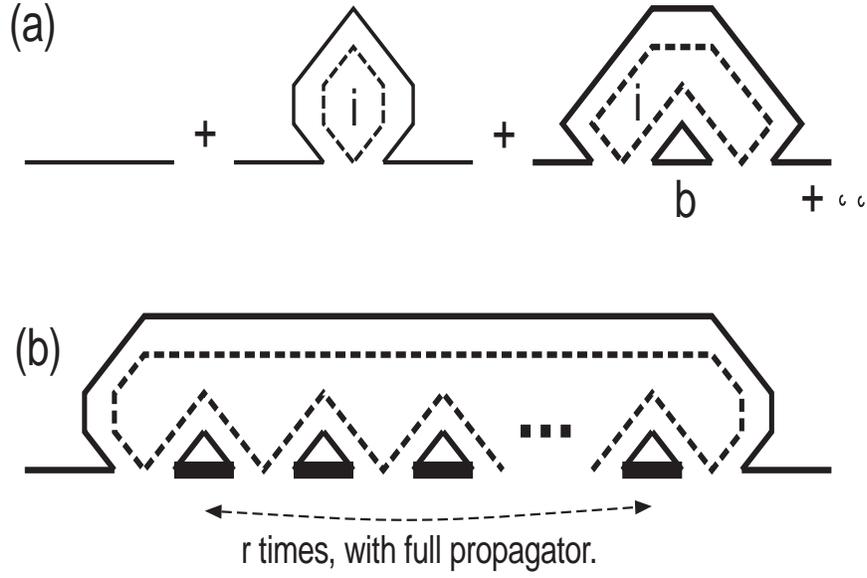,height=3in,width=4.5in}}}
\caption{The diagrams that contribute to the self energy in 
large $p$, $q$ limit. }
\end{figure}

Summing the geometric series in this limit, we obtain 
\begin{equation}
\Sigma (z) = {p\tilde{\sigma}^2 /q \over   1-\tilde {\sigma}^2 G(z)/q}
\end{equation}
\begin{equation}
G(z)={q \over z-\Sigma (z)}= {q \over z- {p\tilde {\sigma}^2 \over q-
\tilde {\sigma}^2G(z)}}
\label{simple}
\end{equation}

The solution of this equation is
\begin{equation}
G(z)={1\over {\tilde \sigma}^2 z}\bigl[-{\tilde \sigma}^2 (p/q -1)
+z \pm \sqrt{4p {\tilde \sigma}^2 -(z-{\tilde \sigma}^2(p/q+1))^2} \bigr
] 
\end{equation}
and
\begin{equation}
\rho(\lambda)={q^{1/2}\over \pi \lambda {\tilde \sigma}}
     \sqrt{(\lambda_{max} ^2 -\lambda ^2)(\lambda ^2- \lambda_{min} ^2)}
\label{sol}
\end{equation}
for $\lambda_{min}  < \lambda < \lambda_{max} $ and zero elsewhere.

\begin{equation}
\lambda_{max,min}={\tilde \sigma}\sqrt {(p/q+1) \pm 2 \sqrt{p/q}}
= \sqrt{2}\sigma \sqrt {(p+q)/2 \pm \sqrt{pq}}
\end{equation}

These results have been obtained by various authors (see for example 
\cite{Denby}).

Generalising our methods, both the saddle point technique and the
perturbative
method, to the following cases is quite easy.

Case 1) 
\begin{equation}
\langle M_{i a} \rangle = M_{ia}^0
\end{equation}
The matrix $M^0$ has singular values $\lambda_{0a}$ where$a=1,\cdots,
q$.

The covariance matrix is given as before by
\begin{equation}
 \langle (M_{i a}-M_{i a}^0)(M_{j b}-M_{j b}^0) \rangle 
=\sigma^2 \delta_{ij} \delta_{ab}
\end{equation}
In this case we obtain 
\begin{equation}
G(z)= \sum_{a=1}^q{1 \over z- \lambda_{0a}^2
-{p {\tilde \sigma}^2\over q-{\tilde \sigma}^2 G(z)}}
=G_0\Bigl (z-{p {\tilde \sigma}^2\over q-{\tilde \sigma}^2 G(z)}\Bigr )
\label{M0}
\end{equation}
where
\begin{equation}
G_0(z)= Tr\Bigl ({1 \over z- M^{0\dagger}M^0}\Bigr )
\end{equation}

In case there are only few non-trivial $\lambda_{0a}$'s, $G(z)$ still
satisfies
a polynomial equation of order two or higher. Denby and Mallows
\cite{Denby}
obtained similar results using a different method.

One of the simple consequences of eqn. \ref{M0} is the following.
Consider
a situation where there are only $r$ non-zero singular values of $M^0$,
each
of which is much bigger than the noise. Let the nonzero SV's be
$\{\lambda_{01},\ldots,\lambda_{0r}\}$. In the limit of zero noise,
\begin{equation}
G(z)=\frac{q-r}{z} + \sum_{a=1}^r \frac{1}{z-\lambda_{0a}^2}
\end{equation}
In presense of finite small $\sigma$, the pole at zero broadens into a
cut
close to the origin, the other cuts develop around the nontrivial
singular
values which are far away from the origin.

Let us try to get the expression of $G(z)$ around the origin. For $z
\sim
{\tilde \sigma}$, ignoring terms of the order 
$({\tilde \sigma}/\lambda_{0a})^2$ for $a\leq r$, we find
\begin{equation}
G(z) \approx \sum_{a=r+1}^q{1 \over z -
{p {\tilde \sigma}^2\over q-{\tilde \sigma}^2 G(z)}}
=\frac{q-r}{ z-{p \sigma^2\over 1- \sigma^2 G(z)}}
\end{equation}
which is just eqn. (\ref{simple}) with $q$ replaced by $q-r$, $\sigma$
(but not ${\tilde \sigma}$) remaining the same. Hence the smallest
$q-r$ singular
values have the same distribution
 as if they have come from a pure noise matrix which is of
a smaller size, namely $p\times(q-r)$. This result is useful
in fitting the formula to the tail of the singular value spectrum 
in a real data matrix, and is used in the fit displayed in Fig.3.
 
Case 2)
 \begin{equation}
\langle M_{ia} \rangle =0
\end{equation}
\begin{equation}
\langle M_{ia} M_{jb} \rangle =C_{ij} D_{ab} 
\end{equation}
$C$ and $D$ being $p\times p$ and $q\times q$ matrices.
Such correlations may arise in imaging data 
when there is spatial inhomogeneity in the variance or
spatial correlations due to filtering of an underlying 
uncorrelated spatial noise distribution, and/or 
when there is temporal filtering of data.

Here we consider
\begin{equation}
\tilde {G} (z) =\Bigl \langle {1 \over z- M^{\dagger}M}\Bigr \rangle
\end{equation}

Note that we did not take the trace, so that $\tilde {G}$ is a matrix. 

We find that $\tilde {G}$ satisfies the equation
\begin{equation}
\tilde {G} (z) ={1 \over z- D Tr{C \over 1-C Tr(D\tilde {G}(z))}}
\end{equation}
To the best of our knowledge, this is a new result. 
To see how to use it, let us consider two special cases.

a) When the noise variance differs from point to point in space.
\begin{equation}
\langle M_{ia} M_{jb} \rangle = \sigma_a^2 \delta_{ij} \delta_{ab}
\end{equation}
In this case,
\begin{equation}
\tilde{G}_{aa}(z)={1 \over 
                z-{\sigma_a^2 \over 1-\sum_b \sigma_b^2
\tilde{G}_{bb}(z)}}
\label{inhomo}
\end{equation}
which is a set of closed equations
for$\tilde{G}_{aa}(z)$,$a=1,\cdots,q$.

b)There are non-trivial temporal correlations, introduced for example by 
linear filtering of an underlying uncorrelated process: 
\begin{equation}
\langle M_{ia} M_{jb} \rangle = C_{ij} \delta_{ab}
\end{equation}
Here
\begin{equation}
G(z)
=Tr\tilde {G}(z)={q \over 
                z-Tr{C \over 1-C G(z)}}
\end{equation}
If the eigenvalues of $C$ are $\sigma_i^2, i=1,\cdots,p$, 
then,
\begin{equation}
G(z)={q \over 
     z-\sum_{i=1}^p{\sigma_i^2 \over
                      1- \sigma_i^2 G(z)}}
\label{manyvar}
\end{equation}

How do we obtain the singular value spectrum from 
Eq.\ref{manyvar}? One way is to rewrite it as
\begin{equation}
z=\frac{q}{G}-\sum_{i=1}^p \frac {1}{G-\frac{1}{\sigma_i^2}}
\label{zofG}
\end{equation}
We want to know $G$ for real $z$. It is useful to  first think of $z$
as a function of $G$ as defined in eqn.\ref{zofG} in the complex $G$
plane.
We now look for level sets of $Im(z(G))$. By tracing the appropriate
branch
of the curve $Im(z(G))=0$, one can solve for $G(z)$ for real $z$.
Taking the imaginary part of the function $G(z)$ thus found gives
the density of singular values. The cumulative density of states, or 
equivalently the sorted singular values, can be found by integrating 
the singular value density. 

Qualitative insight may be gained by realising
that the real and the imaginary parts of z are the two components
of the electric field in a $2$-dimensional electrostatic problem,
with a charge $q$ at the origin, and point charges of strength 
 $-1$ placed at each of the points $(1/\sigma_i^2,0)$, $i=1,\ldots,p$
in the complex $z$ plane. 

In addition to the density of singular values, one can try to compute
the correlations between nearby singular values. It is well known in the 
theory of random matrices that the 
correlation functions of the eigenvalues of a random hermitian
matrix has interesting universal features \cite{Mehta}. 
This is true for eigenvalues chosen from a small enough region, 
so that the eigenvalue density in that region is more or less constant. 
We find that the correlations of the singular values
of a matrix, having  each matrix element $iid$ 
distributed with mean zero and variance $\sigma^2$, 
are in the same universality class as the Gaussian Unitary Ensemble.
The probability density $p(\Delta \lambda)$ of  level spacings 
$\Delta \lambda$ goes as 
$\Delta \lambda^2$ for $\Delta \lambda<< \bar{\Delta \lambda}$.
The probability density of $s=  
\Delta \lambda/ \bar{\Delta \lambda}$ for the Gaussian Unitary
Ensemble is well known in the Random Matrix literature \cite{Mehta}.
It is possible that empirical level-spacing statistics 
can be used as a diagnostic to find out  which singular values 
correspond mostly to `noise' and which correspond mostly to `signal'.

So far, we have discussed what happens to the singular values.
We would also like to estimate the errors made in reconstructing the
matrix
by keeping  a small number terms in the left hand side of \ref{svd}
which correspond to the biggest singular values.
If we keep too few terms, we lose part of the signal. If we keep too
many,
we introduce back the noise. It would be useful to have expressions of
the bias
and the variance of the reconstruction. Unfortunately we do not have a
simple extension of previous techniques to these calculations. Instead
we compute these quanities for small sigma by doing a perturbative
expansion.

Let us go back to case 1), namely when each element of the matrix $M$
is independently distributed with same variance $\sigma^2$ but different
means.
Let
\begin{equation}
M^0_{ia}=<M_{ia}>=\sum_b \lambda_{0b} u^{0b*}_iv^{0b}_a 
\end{equation}
and
\begin{equation}
M_{ia}=\sum_b \lambda_{b} u^{b*}_iv^{b}_a 
\end{equation}
We would like to calculate the mean and the variance  of the
variable  $\hat{M}_{ia}=\sum_{b\in S} \lambda_{b} u^{b*}_iv^{b}_a$
where $S$ is a subset of $\{1,\ldots,q\}$. 

For small $\sigma$,
\begin{eqnarray}
<\hat{M}_{ia}>&=&
\sum_{b\in S} \Bigg[\lambda_{0b} u^{0b*}_iv^{0b}_a
+2\lambda_{0b}^2\sigma^2\sum_{c\neq b}\frac {\lambda_{0c}}
 {(\lambda^{2}_{0b}-\lambda^{2}_{0c})^2}
u^{0c*}_iv^{0c}_a\nonumber\\
& &
-2 \sigma^2 \lambda_{0b} \sum_{c\neq b}
\frac{\lambda^{2}_{0c}}{(\lambda^{2}_{0b}-\lambda^2_{0c})^2}
u^{0b*}_iv^{0b}_a
\Bigg] + o(\sigma^4)
\end{eqnarray}

\begin{eqnarray}
var(\hat{M}_{ia})&=&\sigma^2 \sum_{b\in S}\Bigg[
|u^{0b*}_iv^{0b}_a|^2 + \lambda^{2}_{0b}\sum_{j\neq b}
\frac{(\lambda^{2}_{0b}+\lambda^{2}_{0j})}{(\lambda^{2}_{0b}-\lambda^2_{0j})^2}
 |u^{0j*}_iv^{0b}_a|^2 \nonumber\\
& &
+ \lambda^{2}_{0b}\sum_{c\neq b}
\frac{(\lambda^{2}_{0b}+\lambda^{2}_{0c})}{(\lambda^{2}_{0b}-\lambda^2_{0c})^2}
|u^{0b*}_iv^{0c}_a|^2 
-4\sum_{c\in S, c\neq b}
\frac{(\lambda^{2}_{0b}\lambda^{2}_{0c})}{(\lambda^{2}_{0b}-\lambda^2_{0c})^2}
|u^{0c*}_iv^{0b}_a|^2 
\Bigg] + o(\sigma^4)
\end{eqnarray}

In this expression $j$ runs from $1$ to $p$ with $\lambda_{0j}=0,
\mbox{for}
j>q$. 

To illustrate the utility of these results, we consider the SV
distribution 
obtained from a space-time data set consisting of functional Magnetic 
Resonance Images (fMRI). The experimental details regarding the chosen 
data set can be found in \cite{mitra97MRM}. For our purposes, the 
data constitutes a $1724 \times 550$ matrix. The longer dimension
corresponds
to a subset of the pixels in a $64 \times 64$ spatial image  obtaining
by discarding pixels which have
intensity
below a selected threshhold.
\begin{figure}[h]
\centerline{\hbox{\psfig{figure=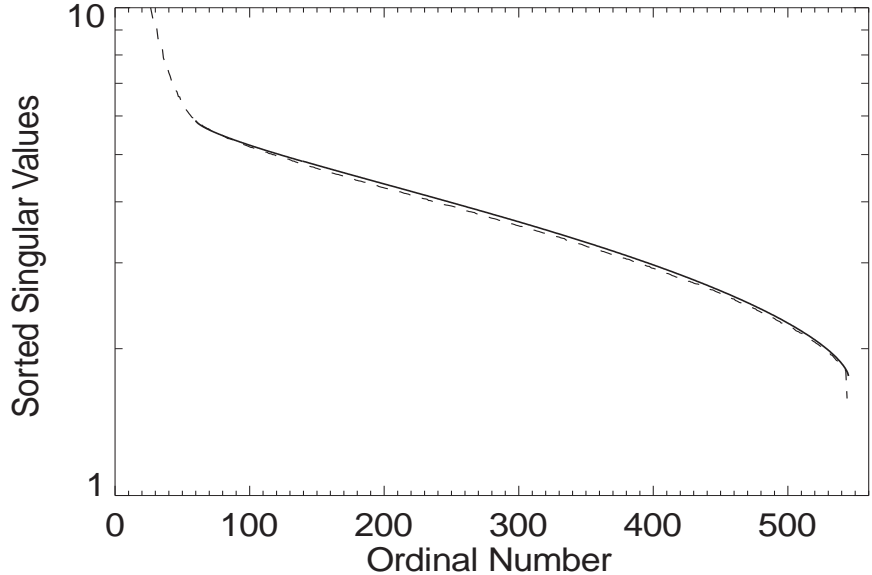,height=3in,width=4.5in}}}
\caption{ Comparison of singular values from a SVD of an fMRI 
data set with the theoretical formula for a noise only matrix. }
\end{figure}
 In Fig.3, the tail of the SV distribution from this data 
is displayed along with a fit to a theoretical curve obtained 
from eqn.(\ref{sol}). The distribution has two adjustable 
parameters. One of them is the variance $\sigma$. A second adjustable 
parameter in the fit is the 
rank of the original matrix, which in this case has been assumed to be
60. We mentioned before that the effect of a few big singular values
coming from the signal on the smaller singular values coming from the
noise
is an effective reduction of the dimensionality of the noise matrix.
We fit the tail to the singular values of a $1724\times (550-60)$
pure-noise matrix. In fact, in the present case the uncorrelated 
noise can be estimated independently, and is therefore not really
a free fitting parameter. We found that the fitted value of $\sigma$
is in close correspondence with the independently estimated 
value of the noise variance (data not shown). 
 
In the example above, the good fit obtained between the theoretical 
curve and the tail in the SV distribution indicates that the noise 
entries in the original data were uniform and uncorrelated. It is easy
to find experimental data where these conditions are violated, for
example
optical measurements of electrical activity in brain tissue
\cite{prechtl97} where the illumination is not fully uniform and the
shot 
noise varies from point to point in space. Alternatively, the data
may be spatially filtered and correlations may be introduced in space
but not in time. Both of these cases produce SV distributions that 
cannot be fit by the procedure described above, but may be understood 
using eqn. (\ref{inhomo}). Details of these applications will be
published 
elsewhere. 

We acknowledge useful discussions with C. L. Mallows. One of
us (AMS) was partly supported by the grant NSF DMR 96-32294.


\begin{references}
\bibitem{Hua} L. K. Hua, Harmonic Analysis of Functions of Several
Complex Variables in the Classical Domains, American Mathematical
Society,
Providence, Rhode Island, 1963.
\bibitem{Denby} L. Denby and C. L. Mallows, ``Computing Sciences and
Statistics:Proceedings of the 23rd Symposium on the Interface",
E. M. Keramidas, Ed.,  54-57, Interface Foundation, Fairfax Station,
VA, 1991.
\bibitem{'tHooft} G. 'tHooft, Nucl. Phys. {\bf B72}, 461 (1974).
\bibitem{Mehta}M. L. Mehta, Random Matrices,Academic Press, New
York,1991.
\bibitem{mitra97MRM} P. P. Mitra, S. Ogawa, X. Hu, K. Ugurbil, Mag. Res.
Med. {\bf 37}, 511 (1997).
\bibitem{prechtl97} J. C. Prechtl, L. B. Cohen, B. Pesaran, P. P. Mitra,
D. Kleinfeld, PNAS {\bf 94}, 7621 (1997).

\end{references}
\end{document}